\documentclass[journal,letterpaper]{IEEEtran}

\usepackage{hyperref}
\usepackage{amssymb}
\usepackage{graphicx}
\usepackage{color}
\usepackage{graphicx}
\usepackage{amsmath}

\usepackage{enumerate}
\usepackage{algorithm}
\usepackage{algpseudocode}
\usepackage{paralist}

\newtheorem{proposition}{Proposition}[section]

\newtheorem{remarks}{Remarks}[section]

\newcommand{\remove}[1]{}

\begin{document}

\title{Neighbor Oblivious and  Finite-State Algorithms for \\ Circumventing Local Minima in \\ Geographic Forwarding}

\author{Santosh Ramachandran, Chandramani Singh, S. V. R. Anand, Malati Hegde, Anurag Kumar, Rajesh Sundaresan
\thanks{The authors are with the Electrical Communications Engineering Department of the Indian Institute of Science, Bangalore, India.}
\thanks{This work was supported in part by a research grant on Wireless Sensor
Networks from DRDO, Government of India, and in part by the Indo-French
Centre for Promotion of Advanced Research~(IFCPAR), Project No. 4000-IT-A.}
\thanks{This work was presented in part at the National Conference on Communications~(NCC)~2010, IIT Madras, Chennai, India.}
}

\maketitle

\begin{abstract}
We propose distributed link reversal algorithms to circumvent
communication voids in geographic routing. We also solve the
attendant problem of integer overflow in these algorithms.
%We solve the imminent problem of integer overflow in link reversal
%algorithms and render a pragmatic solution to circumvent
%communication voids in geographic routing.
These are achieved in two
steps. First, we derive partial and full link reversal
algorithms that do not require one-hop neighbor information,
and convert a destination-disoriented directed acyclic graph~(DAG)
to a destination-oriented DAG.
We embed these algorithms in the framework of Gafni and
Bertsekas~\cite{geo-routing.Bertsekas-etal81-loop-free-algorithms}
in order to establish their termination properties.
We also analyze certain key properties exhibited by our neighbor
oblivious link reversal algorithms, e.g.,
for any two neighbors, their $t$-states are always consecutive integers,
and for any node, its $t$-state size is upper bounded by $log(N)$.
% and define the orientations of routing links among the nodes dynamically based on the state of the network.
In the second step, we resolve the integer overflow problem by
analytically deriving one-bit full link reversal and two-bit partial
link reversal versions of  our neighbor oblivious link reversal
algorithms.

\end{abstract}


\section{Introduction}
\label{sec:introduction}

\subsection{Motivation}
 Consider a wireless sensor network~(WSN) with a single designated
sink node. We shall focus particularly on an example application
where the WSN is used to detect undesirable events. An alarm packet
originating at a node near the location of the alarm event has to be
routed to the sink node.  For such purposes, {\em geographic
routing}~\cite{geo-routing.karp-etal00-GPSR} is a popular protocol
for packet delivery. It is scalable, stateless, and reactive,
without the need for prior route discovery. In this protocol, a node
forwards a packet to another node within its communication
range~(hence, called a \emph{neighbor} node) and
closer to the destination. Ties can be broken arbitrarily, for example,
by using node indices. Such a protocol requires a node with a
packet to be aware of its own geographical location, and that of the sink
and of its neighbors. To each node, the next hop nodes that
are closer to the sink are defined as {\em greedy} neighbors, and
wireless ``links'' are oriented from the nodes to their greedy neighbors.
The resulting routing graph is a {\em directed acyclic graph}~(DAG).

A DAG is said to be {\em destination-oriented} when there is a
directed path in the DAG from any node to the sink. A DAG is {\em
destination-disoriented} if and only if there exists a node other
than the sink that has no outgoing
link~\cite{geo-routing.Bertsekas-etal81-loop-free-algorithms}. The
disadvantaged node with no outgoing links is said to be
{\em stuck}~(as it is unable to forward towards the sink a packet that it receives).
A destination-oriented network under geographic routing may be
rendered destination-disoriented due to various reasons such as node
failures, node removal or sleep-wake cycling.
The failure of geographic routing in the presence of stuck
nodes is commonly referred to as {\em the local minimum
condition}~\cite{geo-routing.fang-etal05-geo-routing-tent}.
Numerous solutions have been
proposed in the literature to pull the network out of a local
minimum condition~(See Section~\ref{lit-survey} for details).
\emph{However, all these solutions require knowledge of one-hop
neighbors and their locations.} Maintenance of one-hop neighbor information, in general,
requires periodic transmissions of {\em keep alive} packets.
\remove{
Nodes
however are duty-cycled to enhance the lifetime of the network. The
intricacies introduced by duty-cycled sensors in the maintenance of
one-hop neighbor information involve several message exchanges
between nodes and their one-hop neighbors, associated access issues
and collision resolution mechanisms. All these can be both time and
energy intensive.
}

We associate each node in the network with a unique numerical value,
henceforth referred to as \emph{state}. A link between a
pair of neighboring nodes is oriented from the node with the higher
state to the node with the lower state. Thus states~(of all the
nodes) determine the routing graph. It is clearly acyclic.
Whenever a node updates its state, it communicates the new state to
its neighbors. Thus all the nodes always have an updated view of the
directions of all their links.\footnote{For communication
between two neighbors, they must have a consistent view of the
direction of the link between them. Thus broadcast of the updated state is
an intrinsic part of all routing algorithms.}
Whenever a node wants to
determine whether it is stuck or not, it broadcasts a {\em hello}
packet containing only its index. All the {\em alive} neighbors
with incoming links from the tagged node acknowledge. If the tagged node does not receive
any acknowledgment until a fixed timeout period, it concludes that
its state is the least among its alive neighbors, i.e., it is stuck. Then, the node
updates its state appropriately to reverse its links.
It also broadcasts the new state
to facilitate its neighbors to update the corresponding link directions.
An update protocol is called {\it neighbor oblivious} if the updating node
does not need to know the {\it exact} values of its neighbors' states.
Neighbor oblivious protocols do not incur the
overhead of neighbor state discovery, and thus save precious
communication time and energy.

Gafni and Bertsekas~\cite{geo-routing.Bertsekas-etal81-loop-free-algorithms}
proposed a general class of distributed {\em link reversal
algorithms} for converting a destination-disoriented DAG into a
destination-oriented DAG. They also described two representative
algorithms, {\em full link reversal} and {\em partial link
reversal}, of their general class. Henceforth, we refer to their
algorithms as {\em GB algorithms}. In the GB algorithms, a stuck node's
update depends on
% in general assume availability of
its one-hop neighbors' states. Thus the GB algorithms
are {\it not neighbor oblivious}.

Our work is motivated by the question: Are there  {\em distributed,
finite-state, neighbor oblivious} protocols that can pull a network out of
its local minimum condition and render it destination-oriented?\footnote{One
simple neighbor oblivious algorithm is to always make a stuck
node increment its state by unity. This algorithm renders the network
destination-oriented but requires a huge number of updates.
In particular, it is neither full link reversal nor partial link reversal.
Recall that each updating node broadcasts its state to determine if it stuck, and then waits
for a timeout period for acknowledgments. Consequently, this simple algorithm
results in significant energy expenditure and delay, and hence, is not desirable.}

\subsection{Related Literature}
\label{lit-survey}
Kranakis et al.~\cite{geo-routing.kranakis1999compass} introduced geographic routing protocols
for planar mobile ad hoc networks, called {\em compass routing} or {\em face routing}.
This technique guarantees delivery in a connected network, but requires
a priori knowledge of full neighborhood.
%Bose et al.~\cite{} proposed
%an improved face routing algorithm, FACE-2.
Karp and Kung~\cite{geo-routing.karp-etal00-GPSR}
presented {\em greedy perimeter stateless routing}~(GPSR) which also ensures
successful routing over planar networks. Kalosha et al.~\cite{geo-routing.kalosha-etal08-Beaconless-Georouting}
addressed a beaconless recovery problem where the local planar
subgraph is constructed on the fly.  Chang et al.~\cite{geo-routing.chang-etal06-RGP}
presented {\em route guiding protocol}~(RGP),
a shortest path routing protocol to bypass voids,
but that also requires communication of current states among neighbors.
Yu et al.~\cite{geo-routing.kim-etal08-hole-detour}
discussed a void bypassing
scheme when both source and sink nodes are mobile. Leong et al.~\cite{geo-routing.leong-etal06-GDSTR}
presented a new geographic routing protocol called
{\em greedy distributed spanning tree routing}~(GDSTR). GDSTR employs convex hulls which
require maintaining topology information.
Casari et al.~\cite{geo-routing.casari-etal06-ALBA} proposed {\em adaptive load-balanced algorithm}~(ALBA),
another greedy forwarding protocol for WSNs.
Some other algorithms developed for mobile adhoc networks include
{\em destination sequenced distance vector}~(DSDV) routing~\cite{geo-routing.perkins1994highly},
{\em wireless routing protocol}~(WRP)~\cite{geo-routing.perkins1994highly}, {\em dynamic
source routing}~(DSR)~\cite{geo-routing.johnson2001dsr} and {\em node elevation
ad hoc routing}~(NEAR)~\cite{geo-routing.arad2009minimizing}.
\remove{
These algorithms provide only a single path for routing between a source destination
pair, but in WRP and DSR, nodes maintain sufficient information to support multipath
routing. A comparison of various adhoc routing algorithms is presented in
Broch et al~\cite{geo-routing.Bertsekas-etal81-loop-free-algorithms}.
}
All the above algorithms require neighbor information at a stuck node,
and some even require
more extensive topology information~(e.g.,~\cite{geo-routing.leong-etal06-GDSTR}).

\remove{
Solutions such as face routing~\cite{geo-routing.karp-etal00-GPSR},
convex hull routing~\cite{geo-routing.leong-etal06-GDSTR}, and link
reversal routing~\cite{geo-routing.Chen-etal06-avoid-void} were
}

Gafni and Bertsekas~\cite{geo-routing.Bertsekas-etal81-loop-free-algorithms}
introduced a general class of link reversal algorithms to maintain routes to the destination.
They also presented
two particular algorithms, the full link reversal algorithm and the partial link
reversal algorithm. The GB algorithms were designed for connected networks. In a
partitioned network, GB algorithms lead to infinite number of state updates without
ever converging. Corson and Ephremides~\cite{geo-routing.corson1995distributed}
presented {\em lightweight mobile routing}~(LMR), a variant of GB link reversal algorithms.
Park and Corson~\cite{geo-routing.park1997highly}
proposed {\em temporally-ordered routing algorithm~(TORA)}  for detecting and dealing with
partitions in the networks. TORA is also an adaptation of GB partial link reversal algorithm
and employs extended states that include {\em current time} and {\em originator id}.
GB link reversal algorithms have also motivated several
{\em leader election} algorithms which are an important building
block for distributed computing, e.g., mutual exclusion
algorithms or group communication protocols.
Malpani et al.~\cite{geo-routing.malpani2000leader}
built a leader election algorithm on the top of TORA for mobile networks.
Ingram et al~\cite{geo-routing.ingram2009asynchronous} proposed a modification
of the algorithm in~\cite{geo-routing.malpani2000leader} that works in an
asynchronous system with arbitrary topology changes.
All these link reversal algorithms employ state variables that
either require infinitesimal precision~(e.g., current time)
or grow unbounded, thus imposing enormous memory requirements. Further, state updates
in these algorithms require frequent information exchanges among neighboring nodes,
and also network wide clock synchronization, thus imposing signification communication overhead.
These drawbacks render the above algorithms unsuitable for large mobile networks with lightweight
mobile nodes. We focus on connected mobile ad hoc networks with single destination
and develop neighbor-oblivious and memory-savvy link reversal algorithms.

Busch and Tithapura~\cite{geo-routing.busch2005analysis} analyzed GB
algorithms~(full and partial link reversal)  to determine the
number of reversals and time until these algorithms converge. Their performance bounds apply
to our algorithms also.

\remove{
The distributed planar graph traversal technique, commonly known as
face routing and presented in~\cite{geo-routing.karp-etal00-GPSR}
and~\cite{geo-routing.Kim-etal05-CLDP}, guarantees delivery if a
path exists. The technique requires a priori knowledge of the full
neighborhood. Kalosha et al.~\cite{geo-routing.kalosha-etal08-Beaconless-Georouting}
addressed a beaconless recovery problem where the local planar
subgraph is constructed on the fly.
%They did not consider the duty-cycling of sensors.
The RGP protocol~\cite{geo-routing.chang-etal06-RGP}
results in a shortest path routing protocol to bypass voids,
but requires communication among neighbors. Yu et al.~\cite{geo-routing.kim-etal08-hole-detour}
discussed a void bypassing
scheme when both source and sink nodes are mobile. Leong et al.~\cite{geo-routing.leong-etal06-GDSTR}
presented a new geographic
routing protocol called GDSTR. GDSTR uses convex hulls which
requires maintaining topology information.
%, an onerous task in a duty-cycled environment.
The ALBA-R protocol~\cite{geo-routing.Petroli-etal07-non-planar-routing-across-dead-ends}
is a nonplanar routing across voids. Routing is based on hierarchy
of colors and is designed to work with ALBA~\cite{geo-routing.casari-etal06-ALBA},
another greedy forwarding protocol for WSN.
Chen et al.~\cite{geo-routing.Chen-etal06-avoid-void} proposed partial link
reversal under the assumption that neighbor information is
available.
}
%In contrast, we propose a
%neighbor oblivious technique to bypass voids.

\subsection{Our Contributions}

We focus on connected mobile ad hoc networks with a single
destination.\footnote{If routing to multiple destinations
is required, for each destination, a logically separate copy of our algorithm should be run.
This limitation is inherent to the class of GB link reversal
algorithms~(see~\cite{geo-routing.Bertsekas-etal81-loop-free-algorithms,
geo-routing.corson1995distributed,geo-routing.park1997highly,
geo-routing.malpani2000leader,geo-routing.ingram2009asynchronous}).}
We propose neighbor oblivious full and partial link reversal~(NOLR)
algorithms in which a stuck node does not need one-hop neighbors' states to
execute its state update. However, as discussed earlier, a node still
has to communicate with its neighbors in order to determine if it
is stuck. But this communication only involves a {\em hello} packet and its acknowledgments,
and thus is ``lightweight''.
% Further, a node also
% needs to communicate its new state to its neighbors
% to facilitate them to update the corresponding link directions.
Then, we embed our NOLR algorithms into the  framework of the GB algorithms.
The embedding ensures that our proposed algorithms render the network
destination-oriented.

In GB and NOLR algorithms, the state spaces are~(countably)
infinite. The reason is that in both the algorithms each node's state grows
without bound with the number of link reversals. The
algorithms therefore cannot be realized in a real operating
environment with only a finite number of bits to represent states, when
repeated link reversals may be encountered. We show that simple
modifications of our NOLR algorithms result in finite-state link
reversal algorithms. At each node, in addition to the initial state,
the full link reversal algorithm requires only
a one-bit dynamic state and the partial link reversal algorithm
requires only a two-bit dynamic state.

Throughout, we assume that new nodes or links are not added to the existing network.
We conclude the paper with a discussion of how addition of new nodes or links affects
our algorithms.

\remove{
Moreover, those are not finite-state
algorithms.
The reason for the infinite number of states in the GB algorithms
and the NOLR algorithms is that each node in the network is
associated with a numerical {\em potential} value, henceforth
referred to as \emph {height},  A stuck
node with no outgoing links initiates a link reversal by increasing
its height so that outgoing links emerge. However, the height
increases exponentially with the number of link reversals. The
algorithms therefore can not be realized in a real operating
environment with finite number of bits to represent heights, when
repeated link reversals may be encountered. We next show that simple
modifications of our NOLR algorithms result in finite-state link
reversal algorithms. The full link reversal algorithm requires only
a one-bit dynamic state and the partial link reversal algorithm
requires only a two-bit dynamic state. In particular, the heights
do not change during the course of the algorithm.
}

\subsection{Organization of the Paper}

 The rest of the paper is organized as follows. In
Section~\ref{sec:GB_overview} we provide an overview of the GB
algorithms. In Section~\ref{sec:full_link_reversal} we discuss full
link reversal. We begin with the NOLR proposal, but with a countably
infinite state space. Then, we make an observation that renders the
NOLR algorithm into a finite-state algorithm without loss of
correctness. In Section~\ref{sec:partial_link_reversal}, we address
partial link reversal, and pass through the same trajectory as for
full-link reversal -- an NOLR algorithm with infinite states
followed by a finite-state version. In Section~\ref{sec:conclusion}
we make some concluding remarks. The appendices contain the detailed
proofs.

\section{Overview of GB Algorithms}
\label{sec:GB_overview}

Consider a WSN with a designated destination node and nondestination nodes
$\{1,2,\dots,N\}$. The nodes are assumed to have static locations.
Two nodes are neighbors if they can directly communicate, and then we say that
there is a {\it link} between them.
 Link reversal schemes can be used in
geographic forwarding by assigning unique states, $a_1,a_2,\dots,a_N$, to
the nodes. The states are totally ordered by a relation $ < $ in the
sense that for any two nodes $i$ and $j$, either $a_i < a_j $ or
$a_j < a_i $, but not both. These states are used in assigning
routing directions to links. The link between a
pair of neighbors is always oriented from the node
with the higher state  to the node with the lower state.
%If the nodes can not communicate, there exists no link between them.

\remove{
These states may correspond to either
hop counts or distances to the destination, and are used in assigning
routing directions to links.
}

\remove{
Gafni and
Bertsekas~\cite{geo-routing.Bertsekas-etal81-loop-free-algorithms}
proposed a general class of distributed link reversal algorithms
that transform a destination-disoriented DAG into a
destination-oriented DAG. They also describe two representative
algorithms, {\em full link reversal} and {\em partial link
reversal}, of their general class.
}

In GB algorithms, the state
associated with a node $i$ is a pair of numbers $(h_i,i)$ for full
reversal and a triplet of numbers $(p_i,h_i,i)$ for partial
reversal, where $h_i$~(called $i$'s {\em height}\footnote{The
heights~($h_i$s) are initialized to either hop counts or distances from
the destination~(evaluated from either actual or virtual locations~\cite{geo-routing.ananth-etal03-NOGEO}),
with the destination's height being
zero.}) and $p_i $ are integers. The ordering $ < $ on the tuples in
each case is the lexicographical ordering.\footnote{For
tuples $a, b$ of the same dimension, $a > b$ iff $a_i > b_i$ where $i$
is the smallest index such that $a_i \neq b_i$.}  For a
node $i$, let $C_i$ denote
%be the set of nodes with which $i$ can communicate~(called $i$'s {\em neighbors}).
$i$'s {\em neighbors}. Also, let $h =
(h_1,\dots,h_N)$ and $p = (p_1,\dots,p_N)$. Then, the forwarding set
of node $i$ can be written as
\[
F_i(h) = \{j \in C_i|~(h_j,j) < (h_i,i)\}
\]
for full reversal, and
\[
F_i(p,h) = \{j \in C_i|~(p_j,h_j,j) < (p_i,h_i,i)\}
\]
for partial reversal. Clearly, node $i$ is {\em stuck} if $F_i(h) =
\emptyset$~(for full reversal), or $F_i(p,h) = \emptyset$~(for
partial reversal). Node $i$, to determine if it is stuck, broadcasts
its state. All its alive neighbors with lower states acknowledge~(Recall
that a few of the neighbors might not be awake due to the sleep-wake cycling
in place). If node $i$ does not receive any acknowledgment until an a priori
fixed timeout, it concludes that its state is the least among its
neighbors, i.e., it is stuck.

The GB algorithms distributively update the states of stuck nodes so that
a destination-oriented DAG is obtained. The algorithms are as follows.

\begingroup
\setlength{\parindent}{0pt}
\paragraph*{Full link reversal}
In this algorithm, a stuck node reverses the direction of all the incoming
links. Node $i$ updates its state as follows.
\endgroup

\begin{algorithm}[h!]
\caption{GB Full link reversal}\label{alg:gbflr}
\begin{algorithmic}[1]
\Loop
\If{$F_i(h) = \emptyset$}
\State $h_i \gets \max\{h_j|~ j \in C_i\}+1$ \label{eq:gbflr-1}
\EndIf
\EndLoop
\end{algorithmic}
\end{algorithm}
\begin{remarks}
Evidently, a node $i$, if stuck, leapfrogs the the heights
of all its neighbors  after an iteration of the above algorithm.
All neighbors thereby enter the forwarding set of node $i$.
\end{remarks}

\begingroup
\setlength{\parindent}{0pt}
\paragraph*{Partial link reversal}
In this algorithm, every node keeps a list of its neighbors that
have already reversed their links to it. If a node is stuck, it
reverses the directions of links to all those neighbors that are not
in the list, and empties the list. If all its neighbors are in the
list, then it reverses the directions of all the incoming links, and
empties the list. Node $i$ updates its state as follows.
\endgroup

\begin{algorithm}[h!]
\caption{GB Partial link reversal}\label{alg:gbplr}
\begin{algorithmic}[1]
\Loop
\If{$F_i(p,h) = \emptyset$}
\State $p_i \gets \min\{p_j|~ j \in C_i\}+1$ \label{eq:gbplr-1}
\If{there exists a $j \in C_i$ with $p_i = p_j$} \label{eq:gbplr-2}
\State $h_i \gets \min\{h_j|~ j \in C_i \mbox{ with } p_i = p_j\} -1$ \label{eq:gbplr-3}
\EndIf
\EndIf
\EndLoop
\end{algorithmic}
\end{algorithm}
\begin{remarks}
All $p_i$s are initialized to $0$. The update rule~(Line~\ref{eq:gbplr-1})
ensures that for neighboring nodes $p_i$s are always adjacent integers.
For a stuck node $i$, the $h_i$ update~(Lines~\ref{eq:gbplr-2}-\ref{eq:gbplr-3})
ensures that, $i$ does not reverse the links to the neighbors that
have updated states since $i$'s last update.
\end{remarks}

Note that all the nodes run Algorithm~\ref{alg:gbflr}~(or Algorithm~\ref{alg:gbplr} in case of partial link reversal)
asynchronously, i.e., their reversals can follow any arbitrary timing and order.
Gafni and Bertsekas~\cite{geo-routing.Bertsekas-etal81-loop-free-algorithms} show
the following properties.
\begin{proposition}
\label{property:GB_termination}
\begin{enumerate}[(a)]
\item Starting from any state $h$, or $(p,h)$, Algorithms~\ref{alg:gbflr}
and~\ref{alg:gbplr} terminate in a finite number of iterations yielding destination
oriented DAGs.
\item Algorithm~\ref{alg:gbflr} results in the same destination-oriented DAG
regardless of the timing and order of reversals. The same holds for
Algorithm~\ref{alg:gbplr}.
\item %\label{property:GB_stuck_node_only_reverse}
Algorithms~\ref{alg:gbflr} and~\ref{alg:gbplr} are such that only those nodes that
do not initially have a greedy path to the destination update their states
at any stage. % at any of the iterations.
\end{enumerate}
\end{proposition}

\begin{remarks}
The updates at a stuck node, in both Algorithms~\ref{alg:gbflr}
and~\ref{alg:gbplr}, depend on knowledge of neighbors'
states~(see Line~\ref{eq:gbflr-1} in Algorithm~\ref{alg:gbflr} and
Lines~\ref{eq:gbplr-1},~\ref{eq:gbplr-2},~\ref{eq:gbplr-3} in
Algorithm~\ref{alg:gbplr}).
%If a stuck node does not know these states,
%it has to establish an alternative link with its
%neighbors and gather these in a reliable fashion.
After each link reversal, the updating node needs to broadcast its new state, and
its neighbors need to gather this information
in a reliable fashion~(e.g., using an error detection scheme).
In the following
sections, we see how to avoid these exchanges, a desired level of
ignorance that we call {\it neighbor obliviousness}.
\end{remarks}
\remove{
Explain the method of response only if greedy here. So
communication  only with those close by.}

\section{Full Link Reversal}
\label{sec:full_link_reversal}

\subsection{Neighbor Oblivious Full Link Reversal}
\label{subsec:nolr_full_reversal}

The main idea may be summarized as follows.
%Consider nodes' updates as in Algorithm~\ref{alg:gbflr}.
Suppose that the algorithm is such
that a node, at any stage,
% and given its own current height,
knows the entire {\em range} of all its neighbors' heights. Then it may
execute a full reversal by raising its height to a value higher
than the maximum in the range. Note that the updating node does not
need to know the exact states of its neighbors, so valuable
communication time and energy are saved.

%These heights are assumed to be distinct\footnote{One may simply append the node index as another component of the state and consider lexicographical ordering, in order to make hop-counts-based initialization result in distinct states.}.

\begingroup
\setlength{\parindent}{0pt}
\paragraph*{Notation} The notation used is listed below for ease of reference.
\endgroup

\begin{itemize}
%\item $N$ is the number of nodes~(excluding the destination).
\item $[N] = \{1,2, \dots, N\} $ is the set of nodes~(or node indices).
%\item $i \in [N] $ is a node index.
\item $t_i \in \mathbb{Z}_+$ is the number of height updates made by
node $i$; this is initialized to $0$ for all $i$.
%\item $t = (t_1, t_2, \ldots, t_N) \in \mathbb{Z}_+^N $. %is the vector of the number of full reversals at each node.
\item $h_i(t_i) \in \mathbb{Z}_{++}$ is the height of node $i$ after $t_i$
updates;\footnote{$\mathbb{Z}_{++} := \mathbb{Z}_+\backslash\{0\}$.} $h_i(0)$ refers to the initial height. The destination's height is $0$.
%\item $z(t)$ is a upper bound on the height of a node after it has executed $t$ full reversals.
\item $a_i = (t_i,h_i(t_i),i)$ is the state of node $i$; $t_i$ is referred to as its
$t$-state.
\item $C_i$ is the set of neighbors of $i$, i.e., those with which $i$ can
directly communicate.
\item $F_i(h) = \{j \in C_i|~(h_j(t_j),j) < (h_i(t_i),i)\}$ is the forwarding set of node $i$, given the heights $h = (h_1(t_1),h_2(t_2),\dots,h_N(t_N))$.
\item $h_{\max} = \max\{h_1(0),\dots ,h_N(0)\}$.
\end{itemize}

The algorithm is simple.
%Each node runs it independently and asynchronously.
Node $i$ updates its state  $a_i$ as follows.
\begin{algorithm}[h]
\caption{Neighbor oblivious full link reversal}\label{alg:noflr}
\begin{algorithmic}[1]
\Loop \If{$F_i(h) = \emptyset$} \State $t_i \gets t_i + 1$
\label{eq:noflr-1} \State $h_i(t_i) \gets  h_i(t_i-1) + h_{\max}$
\label{eqn:flr-h-update} \EndIf \EndLoop
\end{algorithmic}
\end{algorithm}
\begin{remarks}
Node $i$, if stuck, updates its state such that the new height
surpasses the heights of all its neighbors~(see Line~\ref{eq:noflr-1}).
Thus, it reverses all the incoming links.
\end{remarks}

Node $i$ broadcasts a {\em hello} packet to determine if it is stuck.
The lack of feedback~(silence) following a
broadcast suffices to determine if $F_i(h)$ is empty or not.
However, node $i$ does not need to know its neighbors' states to perform updates~(see
Lines~\ref{eq:noflr-1},~\ref{eqn:flr-h-update} in
Algorithm~\ref{alg:noflr}).
Other nodes also independently and
asynchronously execute similar algorithms.
All the nodes broadcast their new states whenever they update.
Timing and order of state
updates can be arbitrary. We now proceed to state and prove some of
the properties of this algorithm.
\begin{proposition}
\label{property:nolr_full_link_reversal}
\begin{enumerate}[(a)]
\item \label{lm:full_eq} The height of a node $i$ in $t$-state $t_i$ is explicitly given by
    \begin{equation*}
    %\label{eq:fbw_fr_h}
        h_i(t_i) =  h_i(0) + t_i h_{\max}.
    \end{equation*}
\item \label{lm:full_z_gr_h} For any node $i$, and $t_i \in \mathbb{Z}_+$, we have $t_i h_{\max} < h_i(t_i) \leq (t_i + 1)h_{\max}$.
%\item \label{lm:full_z_gr_h} For any node $i$, we have $t_i h_{\max} < h_i(t_i) \leq (t_i + 1)h_{\max}$.
%\item \label{lm:full_h_inc} Each update for a stuck node strictly increases its height.
\item \label{lm:full_x_not_y} For any two neighbors $i$ and $j$, and $t_i,t_j \in \mathbb{Z}_+$ we have the following implication
\[
t_i > t_j \Rightarrow h_i(t_i) > h_j(t_j).
\]
\item \label{lm:full_sennd_rx_diff} For any two neighbors $i$ and $j$, at any stage of the algorithm, we have $ 0 \leq \mid t_i - t_j \mid \leq 1$.
\item \label{lm:full_t_bound} For any node $i$, $t_i \leq N$ at any stage of the algorithm.
\end{enumerate}
\end{proposition}

\begin{IEEEproof}
See Appendix~\ref{appendix:nolr_full_link_reversal_properties}.
\end{IEEEproof}

\begin{remarks}
\label{rmrk:noflr_properties}
\begin{enumerate}
\item \label{rmrk:noflr_property1} For any node, the {\it size} of the state~(i.e., the number of bits required to represent the state) grows with the number of state updates. However, Proposition~\ref{property:nolr_full_link_reversal}\eqref{lm:full_t_bound} implies that, for any node, the number of updates is upper bounded by $N$, and hence the $t-$state size is upper bounded by $\log(N)$. Notice that heights are functions of $t$-states~(Proposition~\ref{property:nolr_full_link_reversal}\eqref{lm:full_eq}), and hence need not be stored separately.
\item Proposition~\ref{property:nolr_full_link_reversal}\eqref{lm:full_x_not_y}
implies that the forwarding set of node $i$ can be alternatively
defined as
\[F_i(a) = \{j \in C_i|~a_j < a_i\},\]
where $a = (a_1,\dots,a_N)$ are the nodes' states.
\end{enumerate}
\end{remarks}

\begin{figure}[t]
\centering
    \includegraphics[width=8.0cm,height=2.8cm]{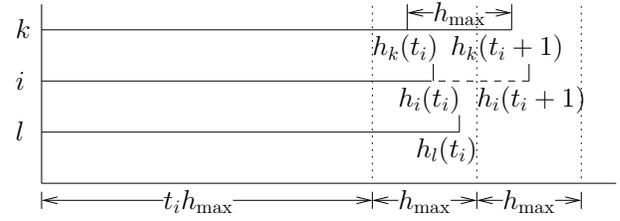}
    \caption{An illustration of Algorithm~\ref{alg:noflr} at a stuck node
    $i$. Note that $t_l = t_i$ while $t_k = t_i +1$. When
node $i$ updates its state, it reverse the links to both $l$ and
$k$.}
    \label{fig:full-reversal}
\end{figure}

\begin{proposition}
\label{property:noflr}
In Algorithm~\ref{alg:noflr}, a stuck node reverses the directions
of all the incoming links.
\end{proposition}
\begin{IEEEproof}
Consider a stuck node $i$. For any node $j \in C_i$, $h_j(t_j) \geq
h_i(t_i)$. So, by virtue of
Propositions~\ref{property:nolr_full_link_reversal}\eqref{lm:full_x_not_y}-\eqref{lm:full_sennd_rx_diff},
we have either $t_j = t_i$ or $t_j = t_i +1$. See
Figure~\ref{fig:full-reversal} for an illustration.
\begin{inparaenum}[(i)]

\noindent
\item Consider $t_j = t_i$. This is the case of node $l$ in Figure~\ref{fig:full-reversal}. In this case, when node $i$ makes an
update, it moves to $t$-state $t_j +1$. Hence the link is from $i$
to $j$ after the update.

\noindent
\item Consider $t_j = t_i + 1$. This is the case of node $k$ in Figure~\ref{fig:full-reversal}. In this case, observe that when node
$j$ updated its $t$-state from $t_i$ to $t_j = t_i+1$, node $i$'s
$t$-state must have been $t_i$. Further, it must have been the case that either
$h_j(t_i) < h_i(t_i)$, or $h_j(t_i) = h_i(t_i)$ and $j < i$. Thus we
have either $h_j(t_j) < h_i(t_i) + h_{\max}$, or $h_j(t_j) =
h_i(t_i) + h_{\max}$ and $j < i$. Hence when node $i$ makes an
update, since $h_i(t_i+1) = h_i(t_i) + h_{\max}$,
the link is now from $i$ to $j$. This concludes the proof.
\end{inparaenum}
\end{IEEEproof}

\begin{proposition}
\label{property:nolr_full_inherit_GB}
Algorithm~\ref{alg:noflr} can
be embedded within the GB algorithms framework. Thus it inherits the
properties in Proposition~\ref{property:GB_termination}.
%and~\ref{property:GB_stuck_node_only_reverse}.
\end{proposition}
\begin{IEEEproof}
See Appendix \ref{appendix:GB_embedding}
\end{IEEEproof}

\remove{
\begin{remarks}
We can use a constant sequence $\{h_{\max},h_{\max},\dots\}$, in
stead of $\{z(0),z(1),\dots\}$, in Algorithm~\ref{alg:noflr}. The
modified algorithm also results in full link reversals for stuck
nodes and can be embedded within the GB algorithms framework.
However, our choice in Algorithm~\ref{alg:noflr} facilitates a
straightforward deduction of finite state version~(see
Section~\ref{sec:tbflr}).
\end{remarks}
}

%Some other crucial properties that will enable a finite-state version of this algorithm are as follows.

\subsection{Two Bits Full Link Reversal}
\label{sec:tbflr} In practice, states are stored using finite
bit-width representations. While the size of the states can depend
on the number of nodes in the network, it should not grow with the
number of iterations of the algorithm. The $t$-states which are the
counts of the number of reversals, though bounded~(see Proposition~\ref{property:nolr_full_link_reversal}\eqref{lm:full_t_bound}),
grow as the algorithm runs. There could be $1000$s of nodes in the network, and in resource limited
nodes in wireless sensor networks, memory is also at a premium.
%the number of link reversals (see Proposition~\ref{property:nolr_full_link_reversal}\eqref{lm:full_eq}),
%and can potentially become too large for realization if repeated link reversals are executed due to node failures.
Therefore, GB and NOLR algorithms need to be modified for implementation in practical
systems.

We now give a modification of Algorithm~\ref{alg:noflr} that uses
only two bits for the $t$-state and does not update heights.
To do this we exploit the fact that, for any two neighbors $i$ and
$j$, the link direction is entirely governed by $t_i,t_j,h_i(0)$ and
$h_j(0)$. More precisely, the link is directed from $i$ to $j$ if
and only if either $t_i > t_j$, or $t_i = t_j$ and $(h_i(0),i) >
(h_j(0),j)$. Thus $t$-states along with the initial heights suffice
to determine link orientations. Moreover, since at any stage $t_i$
and $t_j$ are either the same or adjacent
integers~(Proposition~\ref{property:nolr_full_link_reversal}\eqref{lm:full_sennd_rx_diff}),
we need only two bits to describe their order. Specifically, if we
define, for all $i$,
\[
\tau_i = t_i \mod 4,
\]
and a cyclic ordering
\[
00 < 01 < 10 < 11 < 00
\]
on candidate values of $\tau_i$, we obtain
\[
t_i > t_j \Longleftrightarrow \tau_i > \tau_j.
\]
For node $i$, $\tau_i$ is referred to as its $\tau$-state. Following
the above discussion, we can redefine the forwarding set of node $i$
as
\begin{align*}
F_i(\tau) = \{j \in C_i |&~\tau_j < \tau_i \mbox{ or} \\
                          &~(\tau_j = \tau_i \mbox{ and } (h_j(0),j) < (h_i(0),i))\},
\end{align*}
where $\tau = (\tau_1,\dots,\tau_N)$. In the two bit full link reversal algorithm node
$i$ updates its state as follows.
\begin{algorithm}[h]
\caption{Two bit full link reversal}\label{alg:tbflr}
\begin{algorithmic}[1]
\Loop
\If{$F_i(\tau) = \emptyset$}
\State $\tau_i \gets (\tau_i + 1) \mod 4$ \label{eqn:flr-tau-update}
\EndIf
\EndLoop
\end{algorithmic}
\end{algorithm}

Following are the key properties of this algorithm.
\begin{proposition}
\label{property:tbflr}
\begin{enumerate}[(a)]
\item \label{lm:tb_full}In Algorithm~\ref{alg:tbflr}, a stuck node reverses the
directions of all the incoming links.
\item \label{lm:tbflr_inherit_GB} Algorithm~\ref{alg:tbflr} exhibits the
properties in Proposition~\ref{property:GB_termination}.
%and~\ref{property:GB_stuck_node_only_reverse}.
\end{enumerate}
\end{proposition}
\begin{IEEEproof}
\begin{inparaenum}[(a)]
\item Consider a stuck node $i$. Following Proposition~\ref{property:nolr_full_link_reversal}\eqref{lm:full_sennd_rx_diff}
and the definition of $\tau$-states, for any node $j \in C_i$, we have either
$\tau_j = \tau_i$ or $\tau_j = (\tau_i +1) \mod 4$.
\begin{inparaenum}[(i)]

\noindent
\item Consider $\tau_j = \tau_i$. In this case, when node $i$ makes an
update, it moves to $\tau$-state $(\tau_j +1) \mod 4$ which is greater
than $\tau_j$. Hence the
link is from $i$ to $j$ after the update.

\noindent
\item Consider $\tau_j = (\tau_i +1) \mod 4$. In this case it must be
that $(h_j(0),j) < (h_i(0),i)$; were it not the case, node $j$ at
$\tau$-state $\tau_i$ would not have done an update. Thus when node
$i$ updates its $\tau$-state to $(\tau_i +1) \mod 4 = \tau_j$, the
link is now from $i$ to $j$. This concludes the proof of part~\eqref{lm:tb_full}.
\end{inparaenum}

\noindent
\item Let us consider a network and let all the nodes run Algorithm~\ref{alg:tbflr}.
Also consider another copy of the network~(with the same initial
link orientations) where all the nodes execute
Algorithm~\ref{alg:noflr} as follows. The same node as in the
original network does the first update. Then we are left with the
same set of stuck nodes as in the original network because updates
lead to full link reversals in both the networks. The next update is
also done by the same node as in the original network, thus again
resulting in the same set of stuck nodes. Likewise, subsequent
updates also follow the same timing and order as in the original
network. Since the nodes' updates in the latter network satisfy the
properties in Proposition~\ref{property:GB_termination},
%and~\ref{property:GB_stuck_node_only_reverse}
so do the updates in the original network.
\end{inparaenum}
\end{IEEEproof}

%-------------------------------------
\remove{
 The following section aims to preclude this exponential
growth in height by defining a dynamic link orientation definition
and analytically deriving 1-bit algorithm for full reversal from
neighbor oblivious full link reversal. We exploit the adjacency
property (property \ref{property:nolr_full_link_reversal}
(\ref{lm:full_sennd_rx_diff})), where the $t$ states between two
neighboring nodes are always adjacent to each other, to define the
orientation of links dynamically based on the $t$ state of the
nodes.

        Property \ref{property:nolr_full_link_reversal} (\ref{lm:full_sennd_rx_diff}) highlight the dependence of the $t$
states between two neighboring nodes. By knowing the internal $t$
state of one node, one can determine the $t$ state of its neighbors.
Hence it is possible to define the direction of links between two
neighboring nodes dynamically.
}
\subsection{One Bit Full Link Reversal}
Recall that in full reversal, a stuck node reverses the directions
of all its incoming links. Algorithm~\ref{alg:tbflr} executes this
using initial heights and a two bit state. We now describe a simpler way to achieve this
using initial heights and a single flag bit at each node. More
precisely, with each node $i$, we associate a binary state
$\delta_i$ that is initialized to zero. For any two neighbors $i$
and $j$ with $(h_i(0),i) > (h_j(0),j)$, the corresponding link is
directed from $i$ to $j$ if $\delta_i = \delta_j$, and from $j$ to
$i$ if $\delta_i \neq \delta_j$. In other words, at any stage, the
forwarding set of node $i$ is
\begin{align*}
F_i(\delta) = \{j \in C_i|&~((h_j(0),j) < (h_i(0),i) \mbox{ and } \delta_j = \delta_i) \mbox{ or} \\
&~ ((h_j(0),j) > (h_i(0),i) \mbox{ and } \delta_j \neq \delta_i)\},
\end{align*}
where $\delta = (\delta_1,\dots,\delta_N)$.

We propose the following one bit full link reversal algorithm.
Node $i$ updates its states as follows.
\begin{algorithm}[h]
\caption{One bit full link reversal}\label{alg:obflr}
\begin{algorithmic}[1]
\Loop
\If{$F_i(\delta) = \emptyset$}
\State $\delta_i \gets (\delta_i + 1) \mod 2$ \label{eqn:flr-delta-update}
\EndIf
\EndLoop
\end{algorithmic}
\end{algorithm}
\begin{remarks}
For a stuck node $i$, the updated $\delta$-state is same as the $\delta$-states of neighbors with higher heights but
complements the $\delta$-states  of neighbors with lower heights. Thus, all its links become outgoing.
\end{remarks}
%Note that, unlike Algorithms~\ref{alg:noflr} and~\ref{alg:tbflr}, link orientations now are
Algorithm~\ref{alg:obflr} has similar properties as Algorithm~\ref{alg:tbflr}.
\begin{proposition}
\label{property:obflr}
\begin{enumerate}[(a)]
\item \label{lm:ob_full} In Algorithm~\ref{alg:obflr}, a stuck node reverses the
directions of all the incoming links.
\item \label{lm:obflr_inherit_GB} Algorithm~\ref{alg:obflr} exhibits the
properties in Proposition~\ref{property:GB_termination}.
%and~\ref{property:GB_stuck_node_only_reverse}.
\end{enumerate}
\end{proposition}
\begin{IEEEproof}
\begin{inparaenum}[(a)]
\item Consider a stuck node $i$ and an arbitrary node $j \in C_i$.
Then, either $(h_i(0),i) < (h_j(0),j) \mbox{ and } \delta_i =
\delta_j$, or $(h_  i(0),i) > (h_j(0),j) \mbox{ and } \delta_i \neq
\delta_j$. In either case, when node $i$ flips $\delta_i$, the link
between $i$ and $j$ is reversed.
 \remove{ Node $i$ is stuck. For all $j$,
neighbor of $i$, when $(\delta_i == \delta_j) \Rightarrow h_j >
h_i$, since link is oriented from a node with higher height to a
node with lower height (by definition). When $(\delta_i \neq
\delta_j) \Rightarrow h_j < h_i$, since link is oriented from a node
with lower height to a node with higher height (by definition).

        Node $i$ toggles its $\delta$ value. Now for all $j$, neighbor of $i$, whose $(\delta_i == \delta_j)$ now becomes $\delta_i \neq \delta_j
\Rightarrow $ link is oriented from a node with lower height to a
node with higher height i.e., from $i \rightarrow j$. Similarly, for
all $j$, neighbor of $i$, whose $(\delta_i \neq \delta_j)$ now
becomes $\delta_i == \delta_j \Rightarrow $ link is oriented from a
node with higher height to a node with lower height i.e., from $i
\rightarrow j$.

        Hence a single update of $\delta$ by a stuck node will reverse all its links.

        Here a stuck node toggles its $\delta$ state and reverses all its incoming links.
Thus a full link reversal is realized with a single bit only. }

\noindent
\item The proof is identical to that of Proposition~\ref{property:tbflr}\eqref{lm:tbflr_inherit_GB}.
\end{inparaenum}
\end{IEEEproof}

\section{Partial Link Reversal}
\label{sec:partial_link_reversal}
Recall that the link reversals are intended to yield a destination oriented DAG. However, link
reversals are accompanied by state updates and information exchanges, and can potentially
lead to more nodes being stuck. Thus, a stuck node could execute a partial link reversal~(i.e.,
need not reverse all its incoming links) so that the link graph converges quickly
to a destination oriented graph.
We focus on the partial link reversal scheme proposed by Gafni and
Bertsekas~\cite{geo-routing.Bertsekas-etal81-loop-free-algorithms}~(see Algorithm~\ref{alg:gbplr}).

\subsection{Neighbor Oblivious Partial Link Reversal}
\label{subsec:nolr_partial_link_reversal}

As in neighbor oblivious full link reversal, the algorithm is such
that a node, at any stage,
% and given its own current height,
knows the entire {\em range} of all neighbors' heights but not the
exact values. Then, the node
raises its height to an appropriate value to effect only a partial
link reversal.
\remove{ The partial link reversal at a stuck node's
is executed by raising its height in two steps. The first step
enables reversal of its links that are not previously reversed by
corresponding neighbors. The second step enables reversal of its all
links if all of them are previously reversed by corresponding
neighbors. }
Again, as in Section~\ref{sec:full_link_reversal}, the
updating node does not need to know the exact states of its
neighbors, so valuable communication time and energy are saved.

\begingroup
\setlength{\parindent}{0pt}
\paragraph*{Notation} The new notation is collected below.
\endgroup

\begin{itemize}
%\item $N$ is the number of nodes~(excluding the destination).
%\item $[N] = \{1,2, \dots, N\} $ is the set of nodes~(or, node indices).
%\item $t_i \in \mathbb{Z}_+$ is the number of height updates  made by node $i$; this is initialized to $0$ for all $i$ and includes the updates that do not lead to any reversal.
%\item $\alpha_i \in \{0,1\}$ is the indicator whether a partial reversal was done after the last full reversal; this also is initialized to $0$ for all $i$.
%\item $t = (t_1, t_2, \ldots, t_N) \in \mathbb{Z}_+^N $. %is the vector of the number of full reversals at each node.
%\item $h_i(t_i) \in \mathbb{R}_+$ is the height of node $i$ after $t_i$ updates; $h_i(0)$ refers to the initial height.
%\item $z(t)$ is a upper bound on the height of a node after it has executed $t$ full reversals.
\item $a_i = (t_i,h_i(t_i),(-1)^{t_i}i)$  is the state of node $i$; $t_i$ is referred to as its $t$-state.
%\item $C_i$ is the set of neighbors of $i$.
\item $F_i(h) = \{j \in C_i|~(h_j(t_j),(-1)^{t_j}j) < (h_i(t_i),(-1)^{t_i}i)\}$ is the forwarding set of node $i$ for heights $h = (h_1(t_1),\dots,h_N(t_N))$.
\remove{
\begin{equation*}
F_i(h) = \left\{\begin{array}{ll}
                       , \mbox { if } t_i \mbox{ is even,}\\
                       \{j \in C_i|~(h_j(t_j),-j) < (h_i(t_i),-i)\} , \mbox{ if } t_i \mbox{ is odd.}\end{array} \right.
\end{equation*}
}
%\item $h_{\max} = \max\{h_1(0),\dots h_N(0)\}$.
\item $\{z(0),z(1),\dots\}$ is a sequence satisfying
\begin{equation*}
z(t) = \left\{\begin{array}{ll}
           0 & \mbox{if }  t = 0, \\
           2^{t-1}(2h_{\max} + 1) & \mbox{if } t \geq 1. \end{array} \right.
\end{equation*}
\end{itemize}

In the neighbor oblivious partial link reversal algorithm node
$i$ updates its state as follows.
\begin{algorithm}[h]
\caption{Neighbor oblivious partial link reversal}\label{alg:noplr}
\begin{algorithmic}[1]
\Loop \If{$F_i(h) = \emptyset$} \State $t_i \gets t_i + 1$
\label{eq:noplr-1} \State $h_i(t_i) \gets z(t_i)- h_i(t_i-1)$
\label{eqn:plr-h-update} \EndIf \EndLoop
\end{algorithmic}
\end{algorithm}
\begin{remarks}
Assume that node $i$ is stuck. The height update~(Line~\ref{eqn:plr-h-update}) along with the
definition of sequence $\{z(0),z(1),\dots\}$ ensure that $i$'s updated height
surpasses the heights of those neighbors that have not
updated states since $i$'s last update, but still falls short of the heights of other neighbors.
A similar behavior is ensured by the third components the states~(e.g., $(-1)^{t_i}i$ in $a_i$)
when two neighbors have identical initial heights.
\end{remarks}
\remove{
\begin{figure}[t]
\centering
    \includegraphics[width=6.5cm,height=5cm]{../figures/PAR_FSM}
    \caption{Projection of partial link reversal on state $\alpha_i$ at any node $i$.}
    \label{fig:LR_FSM_projection}
\end{figure}
}

As discussed before, node $i$ broadcasts a {\em hello} packet to determine if
it is stuck. However, it does not need to know its neighbors' states
to perform updates~(see
Lines~\ref{eq:noplr-1},~\ref{eqn:plr-h-update} in
Algorithm~\ref{alg:noplr}).
Also, whenever it updates its state, it broadcasts its new state to facilitate
its neighbors updating the corresponding link directions.
Other nodes also independently and
asynchronously execute similar algorithms. In particular, multiple
nodes can update at the same time. The following properties
of this algorithm are similar to those of Algorithm~\ref{alg:noflr}.
\begin{proposition}
\label{property:nolr_part_link_reversal}
\begin{enumerate}[(a)]
\item \label{lm:part_eq} The height of a node $i$ is explicitly given by
\begin{equation*}
h_i(t_i) =  \left\{\begin{array}{ll}
                       \displaystyle \sum_{l=1}^{t_i/2} z(2l-1) + h_i(0) & \mbox{if } t_i \mbox{ is even,}\\
                       \displaystyle z(1) + \sum_{l=1}^{(t_i-1)/2} z(2l) - h_i(0) & \mbox{if } t_i \mbox{ is odd.}\end{array} \right.
\end{equation*}
\item \label{lm:part_z_gr_h} For any node $i$, and $t_i \in \mathbb{Z}_{++}$, we have $z(t_i-1) < h_i(t_i) < z(t_i)$.
%\item \label{lm:part_lambda_gr_h} For any node $i$, and $t_i \in \mathbb{Z}_+$, we have $h_i(t_i,1) < \lambda(t_i)$.
%\item \label{lm:part_h_inc} Each update for a stuck node strictly increases its height.
\item \label{lm:part_x_not_y} For any two neighbors $i$ and $j$, and $t_i,t_j \in \mathbb{Z}_+$ we have the following implication
\[
t_i > t_j \Rightarrow h_i(t_i) > h_j(t_j).
\]
\item \label{lm:part_sennd_rx_diff} For any two neighbors $i$ and $j$, at any stage of the algorithm, we have $ 0 \leq \mid t_i - t_j \mid \leq 1$.
\item \label{lm:part_t_bound} For any node $i$, $t_i \leq N$ at any stage of the algorithm.
%\item \label{lm:tx=ty} For any node $i,~ j$, $i \neq j$, we have $t_i = t_j,~\alpha_j =1,~\alpha_i=0 \Rightarrow h_j(t_j,1) > h_i(t_i,0)$
\end{enumerate}
\end{proposition}

\begin{IEEEproof}
See Appendix~\ref{appendix:nolr_part_link_reversal_properties}.
\end{IEEEproof}

\begin{remarks}
\begin{enumerate}
\item As in the case of Algorithm~\ref{alg:noflr}, for any node, the number of state updates is upper bounded by $N$, and hence the state size is upper bounded by $\log(N)$.
\item Propositions~\ref{property:nolr_part_link_reversal}\eqref{lm:part_x_not_y}
implies that the forwarding set of node $i$ can be alternatively
defined as
\[F_i(a) = \{j \in C_i|~a_j < a_i\},\]
where $a = (a_1,\dots,a_N)$ are the nodes' states.
\end{enumerate}
\end{remarks}

\begin{figure}[t]
\centering
    \includegraphics[width=8.5cm,height=3.5cm]{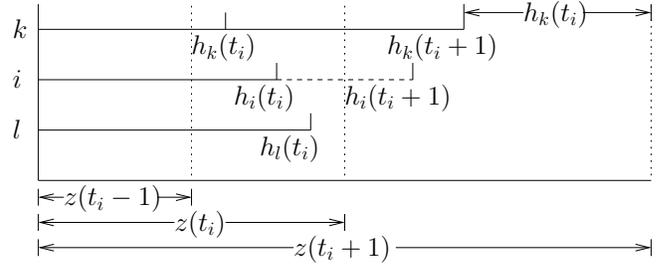}
    \caption{An illustration of Algorithm~\ref{alg:noplr} at a stuck node $i$.
Note that $t_l = t_i$ while $t_k = t_i +1$. Node $k$ has reversed
its link to $i$ after $i$'s last update but node $l$ has not. When
node $i$ updates its state, it reverse the link to $l$ but not the
one to $k$.}
    \label{fig:part-reversal}
\end{figure}

\begin{proposition}
\label{property:noplr} In Algorithm~\ref{alg:noplr}, a stuck node
$i$ reverses the directions of only those of its links that have not
been reversed since $i$'s last update. If every link to node $i$ has
been reversed after $i$'s last update, it performs two successive updates
to reverse the directions of all its links.
\end{proposition}
\begin{IEEEproof}
Since node $i$ is stuck, for any node $j \in C_i$,
\[(h_j(t_j),(-1)^{t_j}j) > (h_i(t_i),(-1)^{t_i}i).\]
By virtue of
Propositions~\ref{property:nolr_part_link_reversal}\eqref{lm:part_x_not_y}-\eqref{lm:part_sennd_rx_diff},
we also have either $t_j = t_i$ or $t_j = t_i +1$. See
Figure~\ref{fig:part-reversal} for an illustration.
\begin{inparaenum}[(i)]

\noindent
\item Consider $t_j = t_i$. This is the case of node $l$ in Figure~\ref{fig:part-reversal}. We claim that node $j$ has not reversed
its link to $i$ since $i$'s last update. If $t_i = 0$, this claim is
trivially valid. If $t_i \geq 1$, we will show that the progression
of updates when both nodes' $t$-states were $t_i-1$ was: node $j$
updated, then node $i$ updated. As a consequence, again, our claim
will be valid. To see the progression of updates, observe that if $h_j(t_j) =
h_i(t_i)$, then $(-1)^{t_j}j > (-1)^{t_i}i$. Thus, by sign flipping,
at $t$-states $t_i - 1 = t_j -1$,  $(-1)^{t_j-1}j < (-1)^{t_i-1}i$.
Also, by the form of the updates at $t_i -1$, $h_j(t_j-1) = h_i(t_i-1)$. So the link was from node $i$ to
node $j$ and it must be $j$ that updated first. On the other hand,
if $h_j(t_j) > h_i(t_i)$, then
\begin{eqnarray*}
h_j(t_j-1) &=& z(t_j) - h_j(t_j) \\
           &<& z(t_i) - h_i(t_i) \\
           &=& h_i(t_i-1).
\end{eqnarray*}
Again we conclude that the link was from $i$ to $j$, and it must be
$j$ that updated first. This establishes the claimed progression of
states.

Continuing with the case, when node $i$ now makes an update, it moves to $t$-state
$t_j + 1$. Hence the link is from $i$ to $j$ after the update.

\noindent
\item Consider $t_j = t_i + 1$. This is the case of node $k$ in Figure~\ref{fig:part-reversal}. We claim that node $j$ has reversed
its link to $i$ after $i$'s last update. Were it not the case, node
$i$'s $t$-state immediately prior to its last update would have been
$t_i - 1 = t_j -2$ which contradicts
Proposition~\ref{property:nolr_part_link_reversal}\eqref{lm:part_sennd_rx_diff}.

Moreover, when node $j$'s $t$-state was $t_j -1 = t_i$, it must have
been the case that
\[(h_j(t_j-1),(-1)^{t_j-1}j) < (h_i(t_i),(-1)^{t_i}i).\]
If $h_j(t_j-1) = h_i(t_i)$, then
$(-1)^{t_j-1}j < (-1)^{t_i}i$. Thus, by sign flipping, at $t$-states $t_i + 1 = t_j$,
$(-1)^{t_j}j > (-1)^{t_i+1}i$. Also, $h_j(t_j) = h_i(t_i+1)$. So,
even after node $i$ makes an updates and moves to $t$-state $t_i+1$,
the link continues to be from $j$ to $i$. If $h_j(t_j-1) <
h_i(t_i)$, then
\begin{eqnarray*}
h_j(t_j) &=& z(t_j) - h_j(t_j-1) \\
           &>& z(t_i + 1) - h_i(t_i) \\
           &=& h_i(t_i+1).
\end{eqnarray*}
Again, even after node $i$ makes an updates and moves to $t$-state
$t_i+1$, the link continues to be from $j$ to $i$. This proves the
first part of the proposition.
\end{inparaenum}

Finally, suppose that every neighbor of node $i$ has reversed its
link to $i$ after $i$'s last update. Then, as shown above,
$t_j = t_i +1$ for all $j \in C_i$. Again as argued above, if node $i$ updates its
state, it does not reverse any of its links, i.e., it is still
stuck. Thus it performs one more update. After this update its
$t$-state is $t_i +2$ which exceeds $t_j$ for all $j \in C_i$. So
all its links are reversed.
\remove{ To prove our first claim, we consider the following two
scenarios separately.
\paragraph*{Case~1~($t_i$ is even)}
Observe that for any node $j \in C_i$
\begin{enumerate}[(i)]
\item  $t_j = t_i$ and $(h_j(0),j) > (h_i(0),i)$ if node $j$ has not reversed its link to $i$ since
$i$'s last update,
\item  $t_j = t_i + 1$ and $(h_j(0),j) < (h_i(0),i)$ if node $j$ has reversed its link to $i$ after
$i$'s last update.
\end{enumerate}
When node $i$ updates its height, it increases $t_i$ by one. In the former case, the updated $t_i$ exceeds $t_j$ and hence
the corresponding link is reversed. In the latter case, the  updated $t_i$ equals $t_j$ and $(-h_j(0),-j) > (-h_i(0),-i)$.
Thus $(h_j(t_j),-j) > (h_i(t_i),-i)$ and the corresponding link is still oriented from $j$ to $i$.

\paragraph*{Case~2~($t_i$ is odd)}
Observe that  for any node $j \in C_i$
\begin{enumerate}[(i)]
\item  $t_j = t_i$ and $(h_j(0),j) < (h_i(0),i)$ if node $j$ has not reversed its link to $i$ since
$i$'s last update,
\item  $t_j = t_i + 1$ and $(h_j(0),j) > (h_i(0),i)$ if node $j$ has reversed its link to $i$ after
$i$'s last update.
\end{enumerate}
When node $i$ updates its height, it increases $t_i$ by one. In the former case, the updated $t_i$ exceeds $t_j$ and hence
the corresponding link is reversed. In the latter case, the  updated $t_i$ equals $t_j$ and $(h_j(0),j) > (h_i(0),i)$.
Thus $(h_j(t_j),j) > (h_i(t_i),i)$ and the corresponding link is still oriented from $j$ to $i$.
}
\end{IEEEproof}

\begin{remarks}
For a stuck node, if all its neighbors have reversed the corresponding links after its last update,
it takes two iteration to reverse all the incoming links. This is unlike Algorithm~\ref{alg:gbplr}
which needs only one iteration.
\end{remarks}

\begin{proposition}
\label{property:nolr_part_inherit_GB}
Algorithm~\ref{alg:noplr} can
be embedded within the GB algorithms framework. Thus it inherits the
properties in Proposition~\ref{property:GB_termination}.
%and~\ref{property:GB_stuck_node_only_reverse}.
\end{proposition}
\begin{IEEEproof}
See Appendix \ref{appendix:GB_embedding}.
\end{IEEEproof}

\subsection{Two-Bit Partial Link Reversal}
In Algorithm~\ref{alg:noplr}, nodes' $t$-states grow as they update.
%exponentially with
%the number of updates~(see
%Proposition~\ref{property:nolr_part_link_reversal}\eqref{lm:part_z_gr_h}).
We now give a modification of Algorithm~\ref{alg:noplr} that uses
only two bits for $t$-state and does not update heights.
To do this we exploit the fact that for any two neighbors $i$ and
$j$, the link direction is entirely governed by $t_i,t_j,h_i(0)$ and
$h_j(0)$.  More precisely, the link is directed from $i$ to $j$ if
and only if either $t_i > t_j$, or $t_i = t_j$ and
$(-1)^{t_i}(h_i(0),i) > (-1)^{t_j}(h_j(0),j)$. Thus $t$-states along
with the initial heights suffice to determine link orientations.
\remove{
\begin{enumerate}
\item $t_i > t_j$
\item $t_i$ is even, $t_i = t_j$  and $(h_i(0),i) > (h_j(0),j)$
\item $t_i$ is odd, $t_i = t_j$,  and $(h_i(0),i) < (h_j(0),j)$
\end{enumerate}
}
Moreover, since at any stage $t_i$ and $t_j$ are either same or adjacent
integers~(Proposition~\ref{property:nolr_part_link_reversal}\eqref{lm:part_sennd_rx_diff}),
we need only two bits to describe their order. Specifically, if we
define $\tau$-states for all the nodes as in Section~\ref{sec:tbflr},
\remove{
\[
\tau_i = t_i \mod 4,
\]
and a cyclic ordering
\[
00 < 01 < 10 < 11 < 00
\]
on candidate values of $\tau_i$, }
we obtain
\[
t_i > t_j \Longleftrightarrow \tau_i > \tau_j.
\]
As before, for node $i$, $\tau_i$ is referred to as its $\tau$-state.
Following the above discussion, we can redefine the forwarding set of node $i$
as
\begin{align*}
F_i(\tau) = \{j \in C_i |&~\tau_j < \tau_i \mbox{ or } (\tau_j = \tau_i\\
                          &~\mbox{ and } (-1)^{\tau_i}(h_i(0),i) > (-1)^{\tau_j}(h_j(0),j))\},
\end{align*}
where $\tau = (\tau_1,\dots,\tau_N)$. We are thus led to the
following two bit version of the partial link reversal algorithm.
Node $i$ updates its states as follows.
\remove{
\begin{align*}
F_i(\tau) = \{j \in &C_i |~\tau_j < \tau_i\\
                  \mbox{or}&~(\tau_i\mbox{ is even}, \tau_i = \tau_j \mbox{ and } (h_i(0),i) > (h_j(0),j)) \\
                  \mbox{or}&~(\tau_i\mbox{ is odd}, \tau_i = \tau_j \mbox{  and } (h_i(0),i) < (h_j(0),j))\}
\end{align*}
}
\begin{algorithm}[h]
\caption{Two bit partial link reversal}\label{alg:tbplr}
\begin{algorithmic}[1]
\Loop \If{$F_i(\tau) = \emptyset$} \State $\tau_i \gets (\tau_i + 1)
\mod 4$ \label{eqn:plr-delta-update} \EndIf \EndLoop
\end{algorithmic}
\end{algorithm}

Following are the key properties of this algorithm.
\begin{proposition}
\label{property:tbplr}
\begin{enumerate}[(a)]
\item \label{lm:tb_part} In Algorithm~\ref{alg:tbplr}, a stuck node
$i$ reverses the directions of only those of its links that have not
been reversed since $i$'s last update. If every link to node $i$ has
been reversed after $i$'s last update, it performs two successive updates
to reverse the directions of all its links.
\item \label{property:tbplr_inherit_GB} Algorithm~\ref{alg:tbplr} exhibits the
properties in Proposition~\ref{property:GB_termination}.
%and~\ref{property:GB_stuck_node_only_reverse}.
\end{enumerate}
\end{proposition}
\begin{IEEEproof}
\begin{inparaenum}[(a)]
\item  Following Proposition~\ref{property:nolr_part_link_reversal}\eqref{lm:part_sennd_rx_diff}
and the definition of $\tau$-states, for any
node $j \in C_i$, we have either $\tau_j = \tau_i$ or $\tau_j =
(\tau_i +1) \mod 4$.
\begin{inparaenum}[(i)]

\noindent
\item Consider $\tau_j = \tau_i$. We claim that node $j$ has not reversed
its link to $i$ since $i$'s last update. If neither $i$ nor $j$ has ever made an update, this claim is
trivially valid. If both of them have made updates,
by Proposition~\ref{property:nolr_part_link_reversal}\eqref{lm:part_sennd_rx_diff},
it cannot be that one of them made two updates without
the other updating. So both must have been at $(\tau_i -1) \mod 4$ at
some point of time. We will show that the progression
of updates when both nodes' $\tau$-states were $\tau_i-1 \mod 4$ was: node $j$
updated, then node $i$ updated. As a consequence, again, our claim
is valid. To see the progression of updates, observe that
\[(-1)^{\tau_j}(h_j(0),j) > (-1)^{\tau_i}(h_i(0),i).\]
Thus, by sign flipping, at the nodes'
immediately prior $\tau$-states, the inequality was in reverse
direction. So the link was from node $i$ to
node $j$ and it must be $j$ that updated first.

Continuing with the case, when node $i$ makes an update, it moves to $\tau$-state
$(\tau_j + 1) \mod 4$. Hence the link is from $i$ to $j$ after the update.

\noindent
\item Consider $\tau_j = (\tau_i + 1) \mod 4$. We claim that node $j$ has reversed
its link to $i$ after $i$'s last update. Were it not the case, node
$i$'s $\tau$-state immediately prior to its last update would have been
$(\tau_i - 1) \mod 4 = (\tau_j -2) \mod 4$ which contradicts
the fact that at any stage $\tau_i$ and $\tau_j$ assume either same or adjacent
values.

Moreover, when node $j$'s $\tau$-state was $(\tau_j -1) \mod 4 = \tau_i$, it must have
been the case that
\[(-1)^{\tau_i}(h_j(0),j) < (-1)^{\tau_i}(h_i(0),i).\]
Thus, by sign flipping, at $\tau$-states $(\tau_i + 1) \mod 4 = \tau_j$,
\[(-1)^{\tau_j}(h_j(0),j) > (-1)^{\tau_j}(h_i(0),i).\]
So, even after node $i$ makes an update and moves to $\tau$-state $(\tau_i+1) \mod 4$,
the link continues to be from $j$ to $i$.
\end{inparaenum}

Finally, suppose that every neighbor of node $i$ has reversed its
link to $i$ after $i$'s last update. Then, by the arguments above,
$\tau_j = (\tau_i +1) \mod 4$ for all $j \in C_i$. Also, if node $i$ updates its
state once, it does not reverse any of its links, i.e., it is still
stuck. Thus it performs one more update. After this update its
$\tau$-state is $(\tau_i +2) \mod 4$ which exceeds $\tau_j$ for all $j \in C_i$. So
all its links are reversed.

\noindent
\item The proof is identical to that of
Proposition~\ref{property:tbflr}\eqref{lm:tbflr_inherit_GB}.
\end{inparaenum}
\remove{
\begin{enumerate}[(a)]
\item Consider a stuck node $i$ and a node $j \in C_i$.
Proposition~\ref{property:nolr_part_link_reversal}\eqref{lm:part_sennd_rx_diff}
implies that $\tau_i$ and $\tau_j$ always assume adjacent values.
More specifically, $(\tau_i + 1) \mod 4 = \tau_j$. Node $i$'s each
update increases  $\tau_i$ by unity, i.e., in two updates node $i$
attains $\tau_i = \tau_j$ and $\tau_i = (\tau_j + 1) \mod 4$
successively. Thus node $i$ reverses the direction of link to node
$j$ in at most two updates. Above arguments apply for all $j \in
C_i$.
\item
}
\remove{
Algorithms~\ref{alg:tbplr} and~\ref{alg:noplr} are related in the same fashion as Algorithms~\ref{alg:tbflr} and~\ref{alg:noflr}.
Thus, the claims can be proved using an argument similar to that in the proof of Proposition~\ref{property:tbflr}.}
%\end{enumerate}
\end{IEEEproof}
\remove{
\begin{remarks}
For a node $i$, the property $|t_i - t_j| \leq 1$~(or, $|\tau_i - \tau_j| \mod 4 \leq 1$ in case of two bit algorithms)
for all $j \in C_i$ holds only under the assumption that new neighbors are
not added to $C_i$, or more generally, new links are not added in the network. Consequently, Propositions~\ref{property:noflr},
~\ref{property:tbflr}\eqref{lm:tb_full},~\ref{property:noplr} and~\ref{property:tbplr}\eqref{lm:tb_part}
hold only under the above assumption.
%; it can take more than one iteration to perform partial or full link reversals if $|t_i - t_j|~ > 1$.
However, all our algorithms continue to exhibit the properties in Proposition~\ref{property:GB_termination}
% and~\ref{property:GB_stuck_node_only_reverse}
even in the general case. Moreover,
Algorithms~\ref{alg:tbflr} and~\ref{alg:tbplr} converge faster than Algorithms~\ref{alg:noflr} and~\ref{alg:noplr}, respectively,
since $| \tau_i - \tau_j |~ \not\geq  4$.
\end{remarks}
}
\section{Conclusion}
\label{sec:conclusion}

We proposed  neighbor oblivious link reversal~(NOLR) schemes to get a destination oriented network
out of the local minimum condition in geographic routing. Our
algorithms fall within the general class of GB
algorithms~\cite{geo-routing.Bertsekas-etal81-loop-free-algorithms}.
We then argued that both the algorithms,
GB and NOLR, may suffer the problem of state storage overflow. This led us to
modify the NOLR algorithms to obtain one bit full
link reversal and two bit partial link reversal algorithms. The
finite state algorithms inherit all the properties of NOLR
algorithms which in turn inherit the properties of GB algorithms,
and are pragmatic link reversal solutions to convert a
destination-disoriented DAG to a destination-oriented DAG.
%We also obtained bounds on the number of updates required to achieve link reversals.

The property
$|t_i - t_j| \leq 1$ at every stage for all pairs of neighboring nodes
is crucial for getting the finite state version of our
NOLR algorithms. If addition of new nodes or links to the existing graph is allowed, this property
could be violated. If full $t$-states~(instead of only $\tau$-states) are maintained,
then since Algorithms~\ref{alg:noflr} and~\ref{alg:noplr} belong to
the class of GB algorithms, they continue to exhibit the properties in Proposition~\ref{property:GB_termination}.
However,  Algorithm~\ref{alg:noflr} does not execute a full link reversal, and
similarly, Algorithm~\ref{alg:noplr} does not execute a partial link reversal.
Furthermore, the finite state algorithms are not robust to addition of new nodes or links because
the newly added nodes may not be able to take up a state consistent with the above property, or the DAG may be burdened
by cycles.

\appendices

\section{Proofs of Propositions~\ref{property:nolr_full_inherit_GB} and~\ref{property:nolr_part_inherit_GB}}
\label{appendix:GB_embedding}

For all $i \in [N]$, let $A_i$ be the set of feasible states of node $i$.
Define $v = (a_1, a_2, \dots, a_N)$. Let $V$ be the set of all such $N$-tuples. For each $v \in V$, let
$S(v) \subset [N]$ denote the set of stuck nodes.
\[
S(v) = \{i \in [N]|~a_j > a_i \mbox{ for all } j \in C_i\}.
\]

We consider iterative algorithms of the form
\[
v \gets \overline{v} \in M(v),
\]
where $M(\cdot)$ is a point-to-set mapping; $M(v) \subset V$ for all $v \in V$. In the following we show that
the proposed neighbor oblivious link reversal algorithms satisfy the assumptions of GB algorithms.

First, we consider Algorithm~\ref{alg:noflr}. Recall that $a_i = (t_i,h_i(t_i),i)$ in this case.
\begingroup
\setlength{\parindent}{0pt}
\paragraph*{(A.1)} Define $g_i:V \rightarrow A_i$, $i = 1,\dots,N$ as
\begin{equation*}
       g_i(v) = \left\{\begin{array}{ll}
                        (t_i + 1, h_i(t_i) + h_{\max},i) & \mbox{if } i \in S(v),\\
                        (t_i, h_i(t_i),i) & \mbox{if } i \notin S(v). \end{array} \right.
\end{equation*}
The set $M(v)$ is then given by
\begin{eqnarray*}
M(v) = \left\{\begin{array}{llll}
                        \{v\}\ \ \ \ \ \ \ \ \ \ \ \ \ \ \ \ \ \ \ \ \ \ \ \ \ \ \ \ \  \ \ \ \ \ \ \ \ \ \mbox{ if }S(v) =\emptyset,\\
                        \{\overline{v} = (\overline{a}_1, \dots, \overline{a}_N)|~\overline{v} \neq v \mbox{ and}\\
                         \ \ \ \ \  \mbox{either } \overline{a}_i = a_i \mbox{ or } \overline{a}_i = g_i(v)\mbox{ for all } i \in [N]\}\\
                         \ \ \ \ \ \ \ \ \ \ \ \ \ \ \ \ \ \ \ \ \ \ \ \ \ \ \ \ \  \ \ \ \ \ \ \ \ \ \ \ \ \ \mbox{ if }S(v)\neq \emptyset.\end{array} \right.
\end{eqnarray*}

\paragraph*{(A.2)} From~(A.1), it is clear that for each $v = (a_1,\dots, a_N)$ and $i = 1,\dots,N$, the functions $g_i(\cdot)$ satisfy
\begin{align*}
g_i(v) > a_i & \mbox{ if } i \in S(v), \\
\mbox{and } g_i(v) = a_i & \mbox{ if } i \notin S(v).
\end{align*}
Furthermore, for each $i = 1,\dots,N$, $g_i(v)$ depends only on $a_i$ and $\{a_j|~j \in C_i\}$;
the latter states determine if $i \in S(v)$ or otherwise.

\paragraph*{(A.3)} Consider a node $i$ and a sequence $\{v^k \} \subset V$ for which $i \in S(v^k)$ for an infinite number of indices $k$.
If $r$ is one of these indices,  $g_i(v^r) - a_i^r \geq (1,h_{\max},0)$, otherwise $g_i(v^r) - a_i^r = 0$.
Hence the sequence
\[
\left\{a_i^0 + \sum_{r = 0}^k [g_i(v^r) - a_i^r]\right\}
\]
is unbounded in $A_i$.
\endgroup
Next, we consider Algorithm~\ref{alg:noplr}. Recall that $a_i = (t_i,h_i(t_i),(-1)^{t_i}i)$
in this case. We define $g_i:V \rightarrow A_i$ as
\begin{equation*}
g_i(v) = \left\{\begin{array}{ll}
                        (t_i + 1, z(t_i + 1)-h_i(t_i),(-1)^{t_i+1}i) \mbox{ if } i \in S(v),\\
                    (t_i,h_i(t_i),(-1)^{t_i}i) \ \ \ \ \ \ \ \ \ \ \ \ \ \ \ \ \ \ \ \ \ \ \mbox{ if } i \notin S(v). \end{array} \right.
\end{equation*}
Again, it is easy to check that Assumptions~(A.1)-(A.3) hold.

Gafni and Bertsekas~\cite{geo-routing.Bertsekas-etal81-loop-free-algorithms} show that if the communication graph is connected and an algorithm satisfies Assumptions~(A.1)-(A.3), then  Proposition~\ref{property:GB_termination}
% and~\ref{property:GB_stuck_node_only_reverse}
holds  for the algorithm. This concludes the proof of
Propositions~\ref{property:nolr_full_inherit_GB} and~\ref{property:nolr_part_inherit_GB}.
\section{Proof of Proposition~\ref{property:nolr_full_link_reversal}}
\label{appendix:nolr_full_link_reversal_properties}
\begin{inparaenum}[(a)]

\noindent
%We now proceed to prove the statements of the proposition.
\item This follows immediately from the height update rule~(Line~\ref{eqn:flr-h-update} in Algorithm~\ref{alg:noflr}).

\noindent
\item This follows from~\eqref{lm:full_eq} and $0 < h_i(0) \leq h_{\max}$.
%This follows from Proposition~\ref{property:nolr_full_link_reversal}\eqref{lm:full_eq}.

\noindent
\item The implication holds because $h_i(t_i) > t_i h_{\max}$ and $h_j(t_j) \leq (t_j+1)h_{\max}$~(see~\eqref{lm:full_z_gr_h}).

\noindent
\item Without loss of generality, assume $t_i \geq t_j$. We claim that $t_i \leq t_j + 1$. We prove the claim via contradiction. Suppose $t_i > t_j + 1$. Node $i$ must have reached this state through $t_j+1$ because $t_i$ is initialized to zero and is incremented by one each time node $i$ updates its state.  When node $i$'s $t$-state was $t_j +1$, from~\eqref{lm:full_x_not_y} $h_i(t_j+1) > h_j(t_j)$, and therefore it had an outgoing link to node $j$. Thus, $i$ would not have updated its $t$-state to $t_j+2$ or higher. This contradicts our supposition, and proves the claim.

\noindent
\item Observe that any one hop neighbor of the destination never updates its heights; it always has an outgoing link
to the destination. Consequently, for any such node, say node $i$, $t_i = 0$ at any stage of the algorithm.
Now, assume that for a node $j$, $t_j > N$ at some stage. Then, there is pair of neighbors $k$ and $l$
such that $\mid t_k - t_l \mid > 2$. But this contradicts part~\eqref{lm:full_sennd_rx_diff}.
Thus, we have the bound $t_i \leq N$ for any node $i$.
\end{inparaenum}

\section{Proof of Proposition~\ref{property:nolr_part_link_reversal}}
\label{appendix:nolr_part_link_reversal_properties}
\begin{inparaenum}[(a)]

\noindent
\item We first obtain a recursion on
$h_i(t_i)$ using the height update rule~(Line~\ref{eqn:plr-h-update} in Algorithm~\ref{alg:noplr}).
For any $t_i \geq 2$,
\begin{eqnarray*}
h_i(t_i) &=& z(t_i) - h_i(t_i-1) \\
         &=& 2z(t_i-1) -(z(t_i-1) - h_i(t_i-2)) \\
         &=& z(t_i-1) + h_i(t_i-2).
\end{eqnarray*}
Successive applications of this recursion leads to expression for the case when $t_i$
is even. If we also use that $h_i(1) = z(1) - h_i(0)$, we get the expression for
the case when $t_i$ is odd.

\noindent
\item We prove the inequalities by induction on $t_i$. For $t_i = 1$,
\[0 < h_i(1) < z(1).\]
Now, assume that $0 < h_i(t_i) < z(t_i)$ for some $t_i \in
\mathbb{Z}_{++}$. From the height update
rule~(Line~\ref{eqn:plr-h-update} in Algorithm~\ref{alg:noplr}),
\begin{eqnarray*}
h_i(t_i+1) &=& z(t_i+1) - h_i(t_i) \\
             &=& 2z(t_i) - h_i(t_i) \\
             &>& z(t_i),
\end{eqnarray*}
where the inequality holds because $h_i(t_i) < z(t_i)$. Also, $0 <
h_i(t_i)$ implies that $h_i(t_i+1) < z(t_i+1)$. This completes the induction, and shows
that the inequalities hold for all $t_i \in  \mathbb{Z}_{++}$.

\noindent
\item The implication holds because $h_j(t_j) < z(t_j)$, $h_i(t_i) >
z(t_i-1)$ and $z(t)$ is increasing in $t$.

\noindent
\item The proof is identical to that of
Proposition~\ref{property:nolr_full_link_reversal}\eqref{lm:full_sennd_rx_diff}.

\noindent
\item The proof is identical to that of
Proposition~\ref{property:nolr_full_link_reversal}\eqref{lm:full_t_bound}.
\end{inparaenum}

\bibliographystyle{IEEEtran}
\bibliography{IEEEabrv,neighbour_agnostic}

\end{document}